\documentclass{aip-cp}

\usepackage[numbers]{natbib}
\usepackage{rotating}
\usepackage{graphicx}

\begin{document}

% Title portion
\title{Resurrecting Quartic and Quadratic inflaton potentials in two-field inflationary model}

\author[aff1]{Suratna Das\corref{cor1}}
%\eaddress{suratna@iitk.ac.in}
\author[aff2]{Girish Kumar Chakravarty}
%\eaddress{}
\author[aff3]{Gaetano Lambiase}
\author[aff2]{Subhendra Mohanty}

\affil[aff1]{Indian Institute of Technology, Kanpur 208016, India}
\affil[aff2]{Physical Research Laboratory, Ahmedabad 380009, India}
\affil[aff3]{Dipartimento di Fisica ``E.R. Caianiello'' Universitá di Salerno,
and INFN-Gruppo Collegato di Salerno, I-84084 Fisciano (Sa), Italy}
\corresp[cor1]{suratna@iitk.ac.in}

\maketitle

\begin{abstract}
After the release of the PLANCK data, it is evident that inflationary paradigm has stood the test of time. Even though, it is difficult to realise inflationary paradigm in a particle physics model as the present observations have ruled out the simplest quartic and quadratic inflationary potentials, which generically arise in particle physics. We would show that such simplest inflationary potentials can evade discrepancies with observations, if the inflaton field is assisted by another scalar during inflation. Moreover, unlike other multifield models, our model yields no isocurvature perturbations and negligible non-Gaussianity, making it more compatible with the present data. Above all, our model can also be realised in the framework of SUGRA.
\end{abstract}

% Head 1
\section{INTRODUCTION}

Inflationary paradigm \cite{Guth:1980zm}, which has become an integrated part of modern cosmology, is a very early phase of the evolutionary history of our universe, during which the universe had expanded exponentially. Such exponential expansion is governed by the slow-roll of a scalar field along its potential, which takes place when the energy density of the scalar field dominates over its kinetic energy. This scalar field is dubbed as the inflaton field, which is a quantum field. The slow-rolling of the inflaton field along its potential during inflation is an important criteria for inflationary dynamics which helps end inflation once the slow-roll seizes, and enter into the standard radiation domination phase of big bang universe. Being a quantum field, the inflaton field generates perturbations both in the matter field and in the background gravitational field. While such primordial perturbations of scalar and tensor types grow during inflation, the vector perturbations decay fast, and thus are of no observational consequences. The slow-roll dynamics of inflation renders a nearly scale invariant power spectrum (two-point correlation function), and negligible non-Gaussianity. This single field slow-roll inflationary model is in very good agreement with observations as the recent PLANCK observation show the scalar spectral tilt is $n_s=0.9666$ \cite{Ade:2015lrj} ($n_s=1$ represents perfect scale invariance) and negligible non-Gaussianity $f_{\rm NL}\sim0$. Presence of more than one scalar field during inflation gives rise to isocurvature perturbations (quantum perturbations generated by presence of other scalar fields during inflation),  which is also being constrained by PLANCK observations very stringently. On the other hand, the ratio of tensor power spectrum amplitude with that of scalar one, known as the tensor-to-scalar ratio, is constrained by PLANCK+BICEP2 observation as $r_{0.05}<0.07\,\,(95\%$ CL) \cite{Array:2015xqh}.

Despite its tremendous success with observations, little is known about the particle physics nature of this inflaton scalar. Also, the slow-roll of the inflaton during inflationary period requires a very flat potential structure. The generic particle physics scalar potentials, quadratic $m^2\phi^2$ and quartic $\lambda\phi^4$, lead to very large tensor-to-scalar ratio, $r\sim0.13$ and $r\sim0.26$ respectively. Thus such simple particle physics scalar potentials are ruled out by observations. We would show here that including another scalar field during inflation in a specific way, would help accommodating quartic and quadratic potentials of the inflaton field and such a model can be realised in a SUGRA model \cite{Chakravarty:2015yho}.  

%=================================================================================================================
\section{THE MODEL AND ITS PREDICTIONS}

We deal with a model where there are two scalar fields : the inflaton field $\phi$ is assisted by a dilaton field $\sigma$. The action of the model can be generically written as 
\begin{eqnarray}
S=\frac12\int d^4x \sqrt{-g}\left[R-\nabla^\mu\sigma\nabla_\mu\sigma-e^{-\gamma\sigma}\nabla^\mu\phi\nabla_\mu\phi-2e^{-\beta\sigma}V(\phi)\right],
\end{eqnarray}
where we would treat $\beta$ and $\gamma$ parameters independently. We define the slow-roll parameters for both the fields in a usual way :
\begin{eqnarray}
\epsilon_\phi&\equiv&\frac{1}{2}\left(\frac{V'(\phi)}{V(\phi)}\right)^2,~~~~~~~~~~~~
\eta_\phi \equiv \frac{V''(\phi)}{V(\phi)},\nonumber\\
\epsilon_\sigma&\equiv&\frac{1}{2}\left(\frac{U_{\sigma}}{U}\right)^2=\frac{\beta^2}{2},~~~~~~~~
\eta_\sigma \equiv \frac{U_{\sigma\sigma}}{U}=\beta^2, \label{slowparameters}
\end{eqnarray}
To ensure the smallness of the slow-roll 
parameters we demand that the Hubble slow-roll parameter $\epsilon_H\equiv-\frac{\dot H}{H^2}\ll1$ during inflation, where $H$ is the Hubble parameter. We notice that 
\begin{eqnarray}
\epsilon_H=\epsilon_\sigma+e^{\gamma\sigma}\epsilon_\phi, 
\label{epsilonH}
\end{eqnarray}
which implies that $\epsilon_\sigma\ll1$ and $e^{\gamma\sigma}\epsilon_\phi\ll1$. This implies that $\gamma\ll1/\beta$.

Analysis of the scalar and tensor perturbations of this model yields the scalar and tensor power spectrum 
\begin{eqnarray}
{\mathcal P}_{\mathcal R}&=&\frac{H^2}{8\pi^2\epsilon_H}, \label{powerR1}\nonumber\\
\mathcal{P}_{\mathcal T}&=&\frac{8H^2}{4\pi^2},
\end{eqnarray}
respectively. The tensor-to-scalar ratio, the scalar spectral index and the isocurvature power spectrum would turn out to be 
\begin{eqnarray}
r&=&16\epsilon_H, \nonumber\\
n_s-1&\simeq& A \left[(2\eta_\phi-6\epsilon_\phi)e^{2\gamma\sigma} - \beta(2\beta+\gamma)e^{\gamma\sigma}\right]
-B \beta^{2},\label{ns}\nonumber\\
{\mathcal P}_{\mathcal S} &=& 
 \frac{H^4}{\pi^2} \frac{\left[\beta e^{-\gamma\sigma}
\dot\phi V(\phi) + \dot\sigma V'(\phi)\right]^2}{e^{(2\beta-\gamma)\sigma}(\dot\sigma^{2}
+ e^{-\gamma\sigma} \dot\phi^2)^3},\label{powerS}
\end{eqnarray}
where $A=\epsilon_\phi/\epsilon_H$  and $B=\epsilon_\sigma/\epsilon_H$. But the isocurvature perturbations vanishes identically upto first order in slow-roll parameters. 
% Figure
\begin{figure}
  \centerline{\includegraphics[width=200pt]{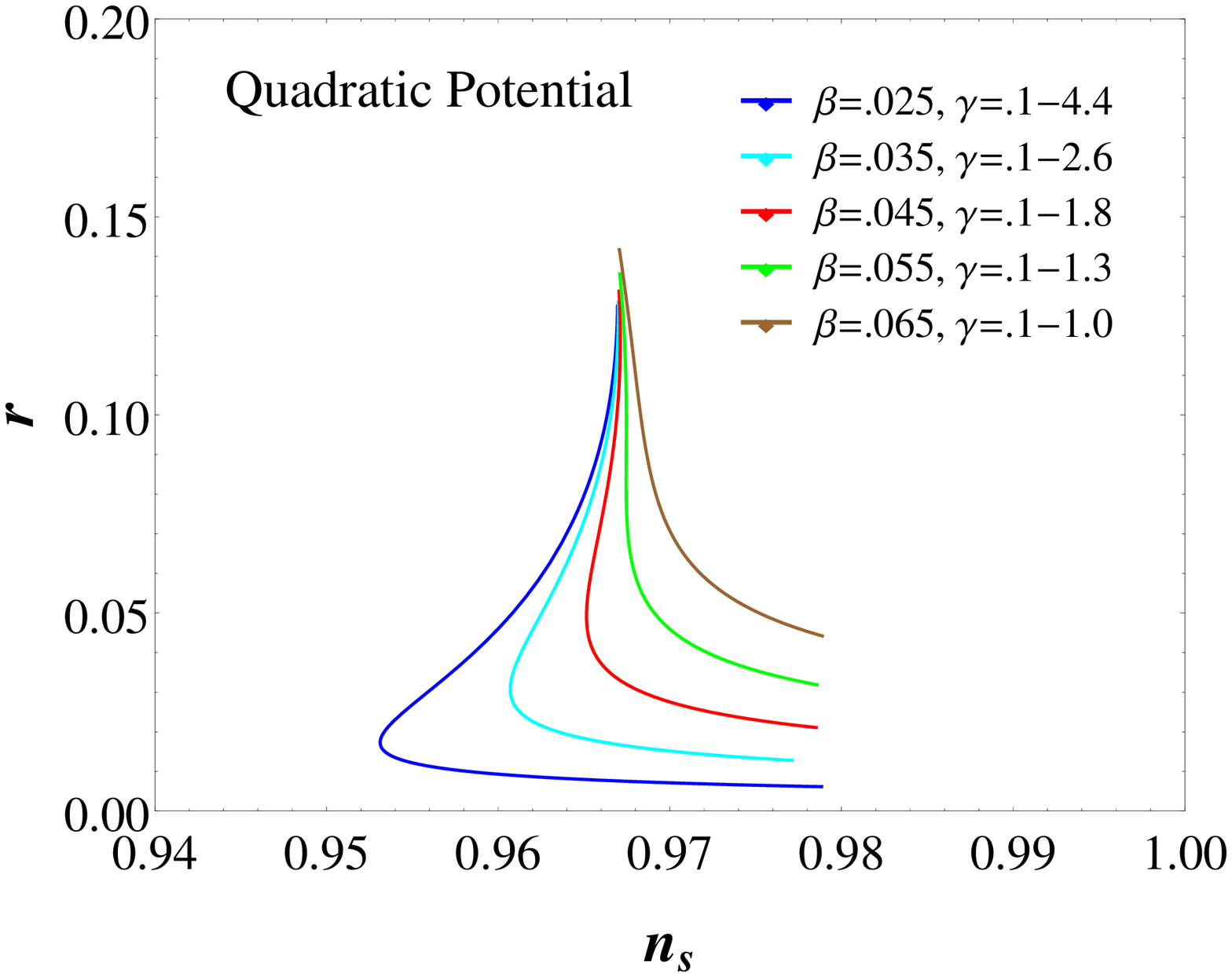}\includegraphics[width=200pt]{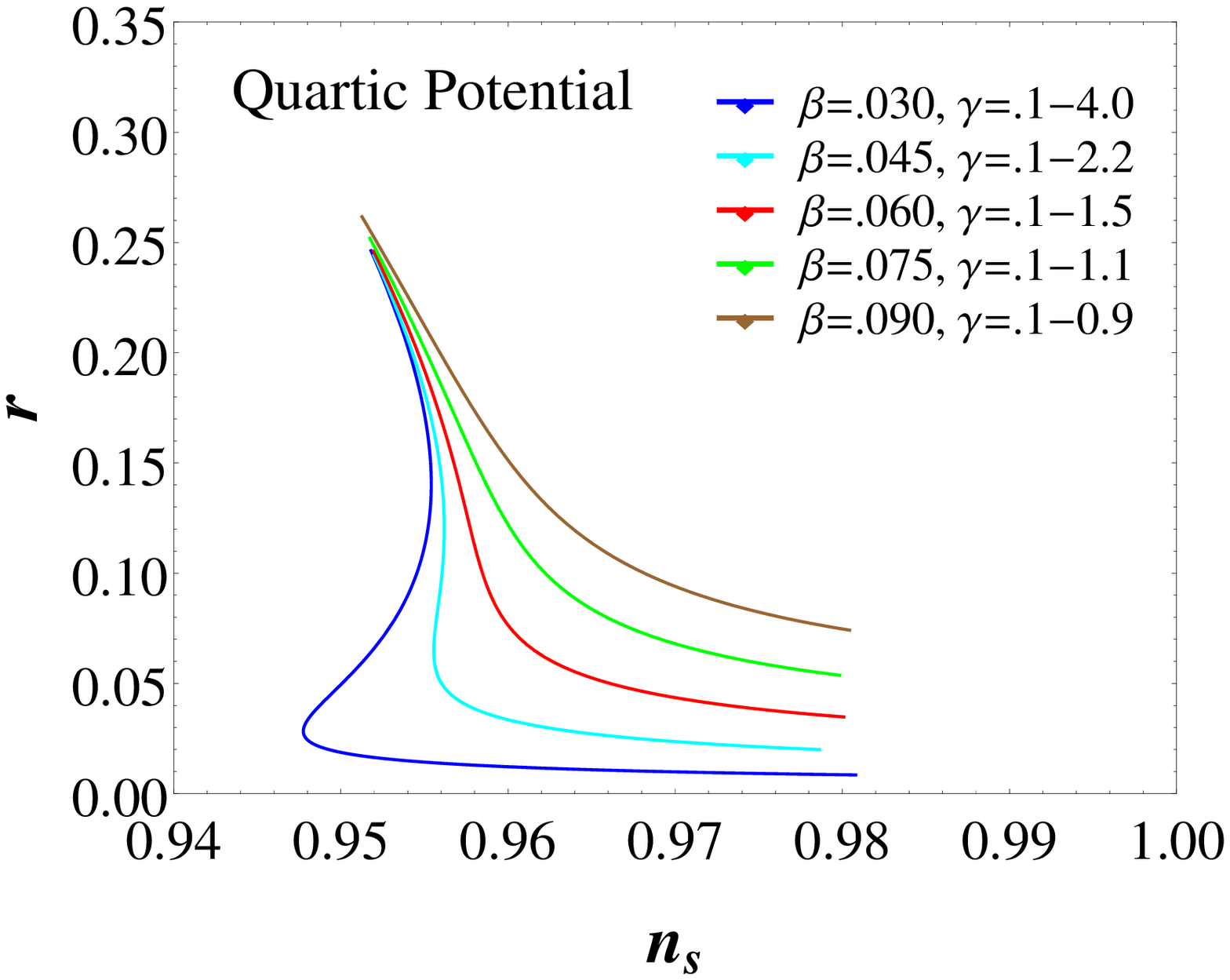}} \label{fig1}
  \caption{$n_s$ vs $r$ plot for quadratic and quartic potential of inflaton. We have taken $\Delta N=60$ and $\sigma_0= 0.1$. In both the figures, the range of values of $\gamma$ increases along the curves from top to bottom. It is also manifest that as the values of $\beta$ and $\gamma$ go to zero, $n_s$ and $r$ values converge to standard slow-roll inflation predictions.
}
\end{figure}

Figure \ref{fig1} shows the $n_s-r$ plot for quadratic and quartic potentials of the inflaton field according to this model. One can see that for certain parameter ranges of $\gamma$ and $\beta$ the model predicts much lower tensor-to-scalar ratio while being in accordance with the scalar spectral index observed value.

%==============================================================================
\section{THE CORRESPONDING NO-SCALE SUGRA MODEL}

\begin{figure}
  \centerline{\includegraphics[width=200pt]{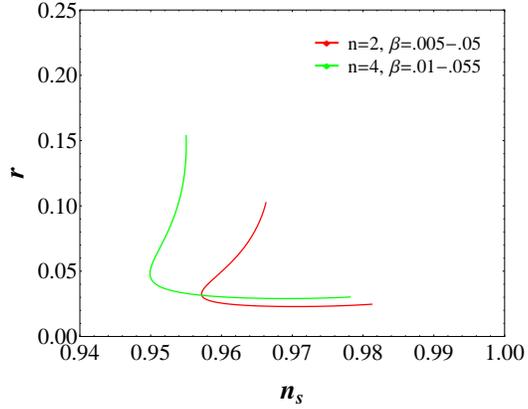}} \label{fig2}
  \caption{$n_s$ vs $r$ for No-scale SUGRA model. We have taken $\gamma=2\sqrt{2/3}$. The range of values of $\beta$ increases along the curves from top to bottom.}
\end{figure}

We consider the potential and kinetic term of the supergravity action as \cite{Ellis:2014gxa}
\begin{eqnarray}
\hskip-0.2cm V=e^{G}\left[\frac{\partial G}{\partial \phi^{i}} K^{i}_{j*} \frac{\partial G}{\partial \phi^{*}_{j}}-3 \right],~~~~~
\mathcal{L}_{K}=K_{i}^{j*} \partial_{\mu}\phi^{i} \partial^{\mu}\phi^{*}_{j}, \label{LVLK}
\end{eqnarray}
where $K^{i}_{j*}$ is the inverse of the K\"ahler metric $K_{i}^{j*} \equiv 
\partial^{2}K / \partial\phi^{i}\partial\phi^{*}_{j}$.
We consider the K\"ahler potential of the following form
\begin{eqnarray}
K=-3 \ln[T+T^{\ast}]+ \frac{b \rho \rho^{\ast}}{(T+T^{\ast})^{\omega}},\label{KP}
\end{eqnarray}
here $T$ is the two component chiral superfield whose real part is the dilaton and imaginary part
is an axion. We identify axion as the inflaton of the model, and $\rho$ is an
additional matter field with modular weight $\omega$. The dynamics of the model ensures that the $\rho$ field decays very fast which leaves us with the inflaton and the dilaton field. Therefore for vanishing $\rho$, the scalar potential and kinetic term (\ref{LVLK}) takes the simple form (the $\omega$ will sustain from the form of $K_i^{\,\,j*}$ even after the $\rho$ field vanishes)
\begin{eqnarray}
V=\frac{\lambda_{m}^{2} T^{m}T^{\ast m}}{b (T+T^{\ast})^{3-\omega}}, ~~~~~~ \mathcal{L}_{K}=
\frac{3 \partial^{\mu}T \partial_{\mu}T^{\ast}}{(T+T^{\ast})^{2}},\label{VLK}
\end{eqnarray}
which upon using the decomposition of $T$ becomes
% \bea
% \mathcal{L}_{K}&=&\frac{1}{2} \partial^{\mu}\sigma \partial_{\mu}\sigma +\frac{1}{2} e^{-\gamma \sigma}
% \partial^{\mu}\phi \partial_{\mu}\phi,\label{LK1} \\
% V&=&\frac{\lambda_{m}^{2} 2^{(\omega-3)}}{b} e^{-\beta\sigma} \left[e^{\gamma\sigma}+\frac{2}{3}\phi^{2}\right]^{m},\label{V1}
% \eea
\begin{eqnarray}
\mathcal{L}_{K}&=&\frac{1}{2} \partial^{\mu}\sigma \partial_{\mu}\sigma +\frac{1}{2} e^{-\gamma \sigma}
\partial^{\mu}\phi \partial_{\mu}\phi,\label{LK1} \\
V&=& \frac{2^{\omega-3}\lambda_{m}^{2}}{b}~ e^{-\beta\sigma} \left[e^{\gamma\sigma}+\frac{2}{3}\phi^{2} \right]^{m},\label{V1}
\end{eqnarray}
where $\gamma=2\sqrt{2/3}\simeq 1.633$ and $\beta=(3-\omega)\sqrt{2/3}$. The prediction of this model is shown in Figure \ref{fig2} which depicts that the model is well in accordance with observations.

\section{CONCLUSION}

We show that if inflaton is assisted by a dilaton field during inflation then the quartic and quadratic potentials can be accommodated within the present observations. Such a two-field inflation model can also be realised in no-scale SUGRA models. 

% Acknowledgement
\section{ACKNOWLEDGMENTS}

Work of S.D. is supported by Department of Science and Technology, Government of India, under the Grant Agreement No. IFA13-PH-77 (INSPIRE Faculty Award). S.D. is thankful to DST-INSPIRE Faculty research grant and INSA for financial support to attend this conference. 

% References

\nocite{*}
%\bibliographystyle{aipnum-cp}%
%\bibliography{two-field-SUGRA-modelNotes}%

\end{document}